\documentclass[a4paper]{VisionStyle}
\usepackage{epsfig}

\newcommand{\aap}{A\&A}
\newcommand{\aaps}{A\&AS}
\newcommand{\apj}{ApJ}
\newcommand{\phrvl}{PhRvL}
\newcommand{\phrc}{PhRvC}
\newcommand{\pasp}{PASP}
\newcommand{\adspr}{AdSpR}
\newcommand{\fluxunit}{${\rm ph/cm^2/s}$}
\newcommand{\tiff}{$^{44}$Ti}

\newcommand{\scff}{$^{44}$Sc}
\newcommand{\msun}{M$_{\sun}$}

\begin{document}

\title{A synoptic spectral study of Cassiopeia-A based on XMM-Newton and 
BeppoSax 
observations.}

\author{Johan\,Bleeker\inst{1} \and Jacco\,Vink\inst{2} \and 
  Kurt\,van der Heyden\inst{1} Dick\,Willingale\inst{3} \and 
  Jelle\,Kaastra\inst{1} \and Martin\,Laming\inst{4} } 

\institute{
 SRON National Institute for Space Research, Sorbonnelaan 2, 3584 CA 
Utrecht, 
the Netherlands
 \and
 Columbia Astrophysics Laboratory, Columbia University, 550 West 120th Street, 
New York, NY 10027, USA (Chandra Fellow)
 \and
 Department of Physics and Astronomy, University of Leicester, University Road, 
Leicester LE1 7RH
 \and
 Naval Research Laboratory, Code 7674L, Washington DC 20375, USA}

\maketitle 

\begin{abstract}
In this paper we present recent image and spectral data of Cas-A
obtained with XMM-Newton and BeppoSAX, which were used to further
constrain the physical characteristics of the source. We analysed the
hard X-ray continuum images between 8 and 15 keV. The data indicate that
the hard  X-ray tail, observed previously, does not originate in
localised regions as demonstrated by the rather flat hardness ratio map
of the 10-15 and 8-10 keV energy bands. This result does not support an
interpretation of the hard X-radiation as synchrotron emission produced
in the primary shock, non-thermal Bremsstrahlung is a plausible
alternative. Moreover, a recent 500 kilosecond deep observation of the
hard X-ray continuum by BeppoSAX has revealed the positive detection of
the $^{44}$Sc nuclear decay lines at 67.9 and 78.4 keV. Appropriate
modelling of the hard X-ray continuum leads to an estimate of the
initial $^{44}$Ti mass of $(0.8-2.5)\times 10^{-4}$ \msun. This implies, by
employing available nucleosynthesis models for type II SNe, a progenitor
mass of $>$30 \msun. We also present a, spatially resolved,
spectral analysis of the thermal emission of Cas-A using detailed
spectral fitting on a 20\arcsec$\times$20\arcsec  pixel scale. This yields maps 
of the
ionisation age, temperature, abundances for Ne, Mg, Si, S, Ar, Ca, Fe
and Ni. The observed elemental abundance ratio patterns can be reconciled  best 
with a type Ib SNe, arising from a 60 \msun\ progenitor (ZAMS) with 
high-mass-loss, which may also explain the relatively high \tiff\ yield. The 
accurate modeling of the
image-resolved emission line spectra has enabled us to derive reliable
Doppler velocities for the bright Si-K, S-K, and Fe-K line complexes.
The combination of radial positions in the plane of the sky and the line
of sight velocities have been used to assess the dynamics of the X-ray
emitting plasma. This indicates a rather asymmetric explosion geometry,
in which the hot Fe-K emitting gas possibly originates from shock-heated
ablated ejecta material, rather than from circumstellar
plasma heated by the primary blast wave.
\end{abstract}

\section{Introduction}

The young galactic supernova remnant (SNR) Cassiopeia A (Cas A) is widely
believed to be the result of a core collapse supernova (SN) observed by 
Flamsteed in 1680 (\cite{jbleeker-B2:ash80}), probably due to the core collapse 
of an early
type Wolf-Rayet star (\cite{jbleeker-B2:fes87}).  Cas A is classified as an
oxygen rich remnant since optical spectroscopic observations
(\cite{jbleeker-B2:ck79}) show the supernova ejecta (in the form of fast moving
knots) to contain mostly oxygen and oxygen burning products such as sulphur,
argon and calcium.  A Wolf-Rayet star progenitor is supported by X-ray data
which, when taking into account that most bremsstrahlung is caused by electrons
interacting with almost completely ionized oxygen, gives a rather low ejecta
mass of $\sim 4$\msun (\cite{jbleeker-B2:vin96}).  At all wavelengths Cas A has
the appearance of a broken shell with a radius varying between 1.6\arcmin\ to
2.5\arcmin.

In this overview of spectral studies of Cas A we highlight the the salient 
features of the XMM-Newton X-ray observatory for SNR research: (i) an 
outstanding spectral grasp brought about by the
large effective collecting area, bandwidth and spectral resolving power of the
X-ray telescopes, (ii) an unprecedented high energy response which allows high
quality X-ray imaging up to 15 keV, (iii) an unique capability for high
resolution (R$>$100) spectroscopy of moderately extended sources. The medium 
resolution spatially resolved spectroscopy of Cas A with the solid
state (\textit{EPIC-MOS} and \textit{EPIC-PN}) X-ray camera's of XMM-Newton 
(\cite{jbleeker-B2:wil02})
provides a powerful tool to study in detail the dynamics, the distribution and
the physical properties of the (reverse) shock heated plasma, while the
bandwidth (up to 15 keV) offers an unique opportunity to assess the origin of
the hard X-ray continuum (\cite{jbleeker-B2:ble01}). We also report on a recent 
deep exposure of Cas A with the BeppoSAX observatory. 
This observation gives a reliable
estimate of the nuclear de-excitiation lines of $^{44}$Sc at 67.9 keV and 78.4 
keV,
produced in the decay chain 
$^{44}$Ti$\rightarrow$$^{44}$Sc$\rightarrow$$^{44}$Ca.
The detection of these lines and the abundance patterns obtained from the 
\textit{EPIC-MOS} analysis 
provides independant measures of the zero-age-main-sequence (ZAMS) progenitor 
mass.

\section{Origin of the hard X-ray continuum}

One of the outstanding problems in the study of Cas A is the origin of a
recently discovered hard X-ray continuum out to 100 keV in the spectrum of this 
remnant (e.g. \cite{jbleeker-B2:the96}; \cite{jbleeker-B2:fav97}; 
\cite{jbleeker-B2:all97}). From a theoretical point of view both synchrotron
emission from shock accelerated electrons and non-thermal Bremsstrahlung from
electrons accelerated into a tail of the Maxwellian velocity distribution could
be possible explanations for the hard continuum. Previous hard X-ray imaging
observations, for example using BeppoSAX (\cite{jbleeker-B2:vin99}), indicated
that the hard continuum radiation originated predominantly in the West region 
of
the remnant, however this result was based on deconvolved images from a
typically 1-arcminute resolution telescope with moderate effective area. The
combination of large collecting area in the energy band 4.0 to 15.0 keV and the
angular resolution of a few arcseconds of XMM-Newton provides a unique
opportunity to search for the distribution and origin of this hard ``tail''.

\begin{figure} 
\centerline{
\resizebox{9.5cm}{!}
{\includegraphics{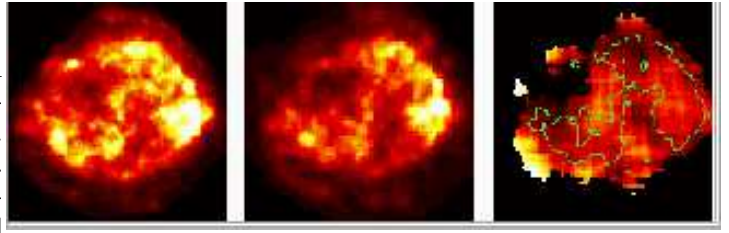}} }
\caption[]{Continuum maps of Cas A in the 4-6 keV band (left) and 
the 8-15 keV band (centre). The image
on the right displays the hardness ratio between the 10-15 keV and the 8-10 keV
emission (\cite{jbleeker-B2:ble01}).} 
\label{fig1} 
\end{figure}

XMM-Newton clearly detects the
hard X-ray tail from the remnant but the hard X-ray image (8-15 keV) displays a
remarkably similar brightness distribution to the softer thermal components,
which presumably dominate the 4-6 keV continuum emission as shown in 
figure~\ref{fig1}
(centre and left respectively). 
Spectral fits to the Cas A data (\cite{jbleeker-B2:ble01}) indicate that the 
ratio
between a powerlaw and thermal Bremsstrahlung components runs from about 1 in 
the 8-10 keV
interval to about 3 in the 10-15 keV band.
A hardness ratio map (10-15)/(8-10) is shown in figure~\ref{fig1}
(right). Although this map shows significant changes in hardness over the
remnant, all the bright features within the green contour have a remarkably
similar spectrum above 8~keV. Therefore the hard X-ray image and the hardness
ratio indicate clearly that this flux does not predominate in a few localised
regions, as would be expected from straightforward synchrotron emission models,
e.g. a limb brightened shell commensurate with the radio morphology.
Apparently the hard X-ray emission pervades the whole remnant with a
distribution similar to the thermal continuum emission. In conclusion, while
the spectral form of the non-thermal high-energy ``tail'' is not inconsistent
with a simple model of synchrotron emission from SNRs 
(\cite{jbleeker-B2:rey98}), the morphology of the hard X-ray image does not 
support this
interpretation in any way. An alternative explanation for the observation of
hard X-ray tails in the spectra of supernova remnants is the presence of
non-thermal Bremsstrahlung generated by a population of suprathermal electrons
(\cite{jbleeker-B2:asv89}; \cite{jbleeker-B2:lam00}).

\section{\scff\ line emission and the hard X-ray continuum}
\label{}

\begin{figure} 
\centerline{
\resizebox{7.5cm}{!}
{\includegraphics{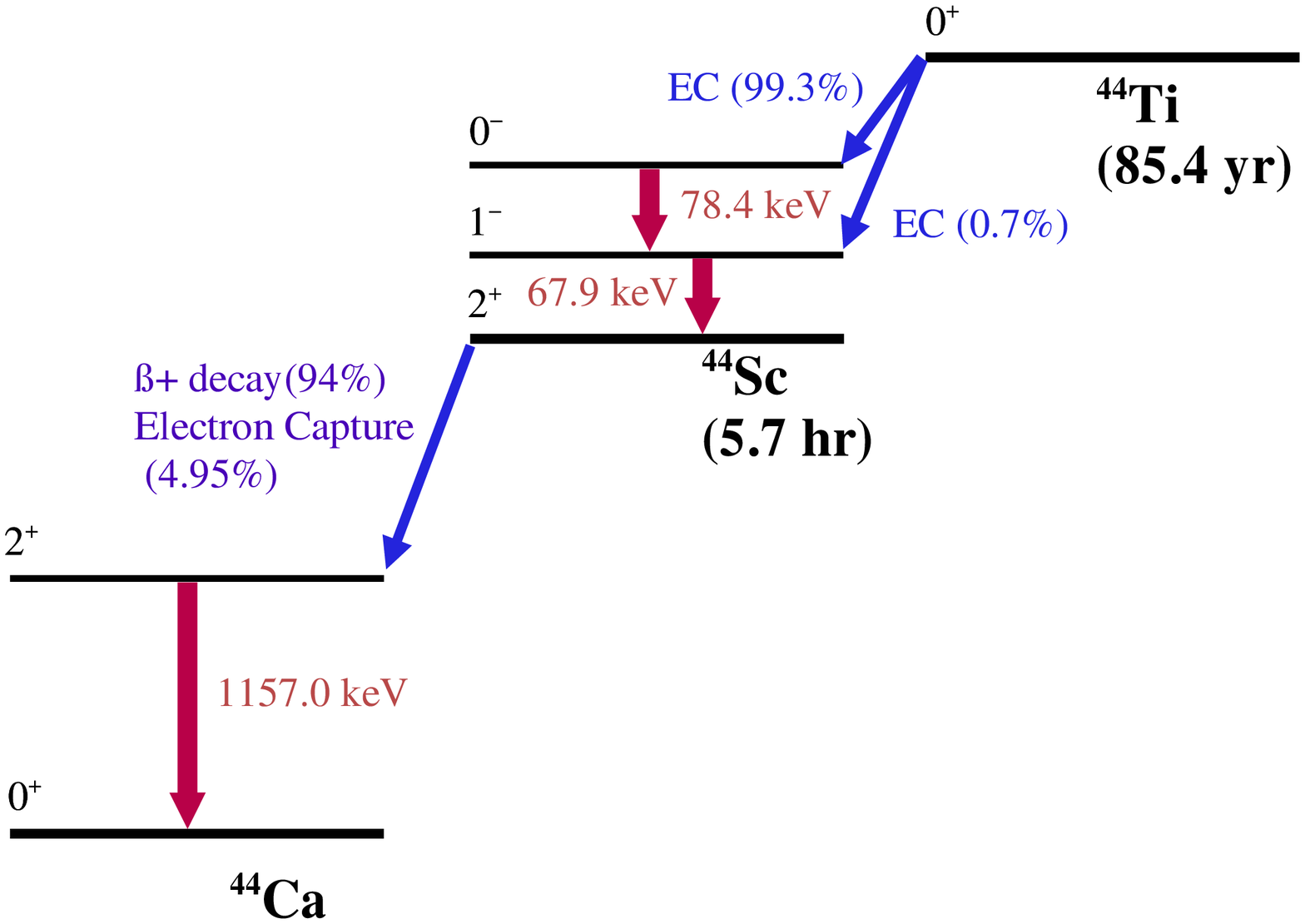}} } 
\caption[]{Decay scheme of \tiff.}
\label{fig2} 
\end{figure}

Observing the nuclear decay emission associated with \tiff\ (see figure 
\ref{fig2}) is probably
the most direct way of observing the abundance of an element for a young
supernova remnant. It does not depend on plasma properties, 
and temperature, except in
the very unlikely case that \tiff\ is almost completely ionized,
which requires too high a temperature (\cite{jbleeker-B2:lam01b}).
If we know the distance to Cas A and the decay time of \tiff,
which has recently been measured accurately to be 
 $85.4 \pm 0.9$~yr 
(\cite{jbleeker-B2:ahm98,jbleeker-B2:goe98,jbleeker-B2:nor98}),
calculation of the initial \tiff\ mass is straightforward. 

Compared to other abundantly produced radio-active elements it has 
a relatively long decay time, which makes this the primary
energy source for late time light curves of supernovae; 
in SN1987A this occurred after $\sim$2000~days, 
when most of the  $^{56}$Co had decayed (\cite{jbleeker-B2:dt98}).
Its main importance is, however, the diagnostics it provides for
Type II supernova explosions.
\tiff\ is thought to be synthesized during explosive Si burning, as the
expanding plasma cools in the presence of sufficiently abundant, 
free $\alpha$ particles,
which is the case at relatively low densities 
(\cite{jbleeker-B2:dt98};\cite{jbleeker-B2:the98}). 
This process is referred to as
$\alpha$-rich freeze out. 
As a consequence, its synthesis is very sensitive to the kinetic energy
and asymetries of the explosion (\cite{jbleeker-B2:nag98}).
Moreover, the amount of \tiff\ ejected depends on the position of the 
mass-cut, the border that divides the 
material falling back on the neutron star and that being ejected.
Also the mass-cut depends both on the energy of the explosion, and the mass
of the progenitor; the expansion velocity is larger 
for less massive progenitors, which allows more core material to escape the
gravitational potential of the neutron star or black hole.

The detection of 1157~keV line emission by CGRO-COMPTEL 
(\cite{jbleeker-B2:iyu94}),
was the first time that the presence of \tiff\ in supernova remnants was 
actually observed. However, subsequent searches with hard X-ray detectors 
for the \scff\ line emission at 68~keV and 78~keV did not confirm the
presence of \tiff\ at the expected flux level.
A deep (500 ks) observation of Cas A with BeppoSAX, combined with archival data, 
has provided us with a net Phoswhich Detector System (PDS) spectrum
of 311~ks (the PDS efficiency is 50\%, as always one of the two
rocking collimators is measuring a 210\arcmin\ off source background).

As reported by \cite*{jbleeker-B2:vin01} the deep observation resulted in a 
larger than
3.4$\sigma$ detection of the \scff\ lines, 
each with a flux of $(2.1 \pm 0.7)\times 10^{-5}$ \fluxunit (90\% error).
This compares well with a more recent CGRO-COMPTEL measurement of
$(3.3 \pm 0.6)\times 10^{-5}$ \fluxunit\ (\cite{jbleeker-B2:iyu97}).
Combining the two measurements suggest an 
initial \tiff\ mass of $1.2\times 10^{-4}$~\msun,
with a 90\% uncertainty range of $(0.8 - 2.5)\times 10^{-4}$~\msun.

\begin{figure} 
\centerline{
\resizebox{8.5cm}{!}
{\includegraphics[angle=-90]{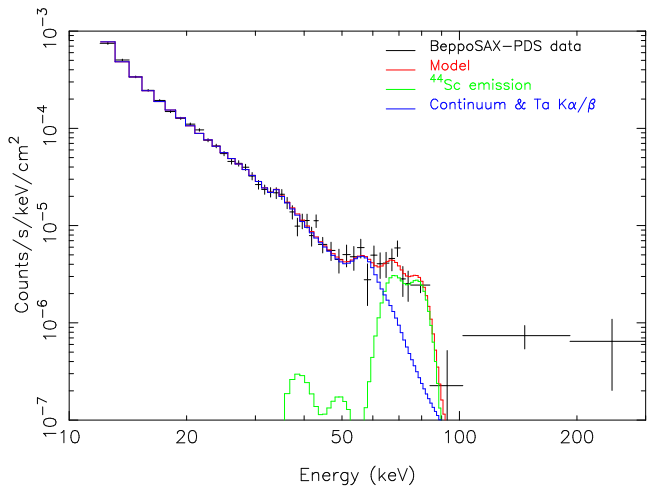}} } 
\caption[]{Cas A's hard X-ray spectrum as observed by the BeppoSAX PDS 
instrument.
The continuum model is that of \cite*{jbleeker-B2:lam01a}. 
Observed countrates have been divided by the effective area to provide an 
approximate photon flux density.} 
\label{fig3} 
\end{figure}

The major uncertainty in measuring the \scff\ flux is the contribution
of the hard X-ray continuum to the spectrum around 70~keV,
due to the uncertain
nature of the hard X-ray emission already addressed in the previous paragraph.
\cite*{jbleeker-B2:vin01} employed various models, including a hard X-ray 
synchrotron model. Here we expand on that measurement by using a model spectrum 
that includes a non-thermal bremsstrahlung continuum caused by electrons 
accelerated
by lower hybrid waves, a process that is likely to be most efficient
in the presence of heavy nuclei, which suggest that the hard X-ray emission
originates from material heated by the reverse shock 
(\cite{jbleeker-B2:lam00,jbleeker-B2:lam01a}).
Note that this model is different from that proposed by 
\cite*{jbleeker-B2:asv89},
who suggest that the non-thermal electrons are the non-relativistic part
of the electron cosmic ray population.

The bremsstrahlung model was calculated using fully relativistic gaunt factors
and, apart from the non-thermal continuum, a 3.5~keV 
thermal component was included and Ta K$\alpha$\ and K$\beta$ line emission,
originating in the PDS collimator, which consists of tantalum.

The best fit non-thermal model has $\chi^2/{\rm d.o.f} = 273/252$ and
suggests a maximum electron acceleration of $\sim$95~keV 
(figure \ref{fig3}). As the continuum drops near the \scff\ emission
the inferred line contribution is larger 
(($3.2\pm 0.4)\times 10^{-4}$ \fluxunit). This is very close to the latest 
COMPTEL
results, and implies an initial \tiff\ mass of $1.5\times 10^{-4}$~\msun.
However, the model does not fit the data above 100~keV well.
This may not be too much of a problem, since the statistics above 100~keV is 
poor and the signal may still be contaminated by residual background.
However, assuming that most of the hard X-ray emission is indeed caused by
electrons accelerated by lower hybrid waves, additional components,
such as the one proposed by \cite*{jbleeker-B2:asv89}, or inverse Compton 
radiation
(see e.g. \cite*{jbleeker-B2:all97}), are likely to be present at some level.

What is the implication of the observed amount of \tiff\ ?
Comparing it to the predicted \tiff\ mass
(\cite{jbleeker-B2:wv95,jbleeker-B2:tim96}), it seems that the actual \tiff\ 
mass is higher than expected, except perhaps models S30B, S35C and S40C 
of \cite*{jbleeker-B2:wv95}, which model
explosions with more than $2\times 10^{51}$~ergs of kinetic energy. 
Apart from energetics, the \tiff\ results for Cas A may be explained by 
pre-supernova mass loss, as is expected to be the case for type Ib SNe 
(\cite{jbleeker-B2:tim96}). Moreover, an asymmetric
explosion would also contribute to a larger \tiff\ abundance. In fact, from the 
following sections it will become clear that the
Cas A explosion indeed displays a high degree of asymmetry. Less certain is 
whether Cas A may also have been a much more energetic explosion.

\section{Plasma physics and dynamics}

\begin{figure}[!htb]
\centering
\rotatebox{270}{\includegraphics[width=6cm]{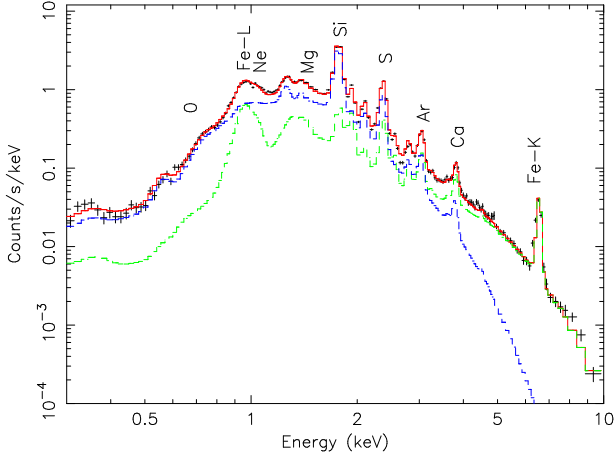}}
\caption{An example of a spectral fit within a single 
$20{\arcsec} \times 20{\arcsec}$ pixel - cool component in blue, hot component
in green and full model in red.}
\label{fig4}
\end{figure}

The brightness of Cas A coupled to a long exposure of XMM-Newton's solid state
cameras makes it possible to do a detailed X-ray spectral analysis with an
angular resolution of 20 arcseconds over the the full extent of the remnant.
The energy resolution, gain stability and gain uniformity of the 
\textit{EPIC-MOS}-cameras
allows significant detection of emission line energy shifts of order 1 eV or
greater for prominent emission lines like Si-K, S-K and Fe-K. Proper modeling
of these line blends with the aid of broad band spectral fitting, taking into
account the non-equilibrium ionisation (NEI) balance, allows an assessment of
Doppler shifts and abundance variations of the X-ray emitting material across
the face of the remnant with adequate spatial resolving power. Two NEI
components for the thermal emission were used as a minimum for representative
spectral modeling. In addition the model accounts for two separate redshifts
(one for each plasma component) and the amount of foreground absorption as free
parameters. Figure \ref{fig4} shows a typical spectral fit.

\subsection{Temperature and Ionisation Age}

\begin{figure}
\centerline{
\resizebox{7cm}{!}{\includegraphics{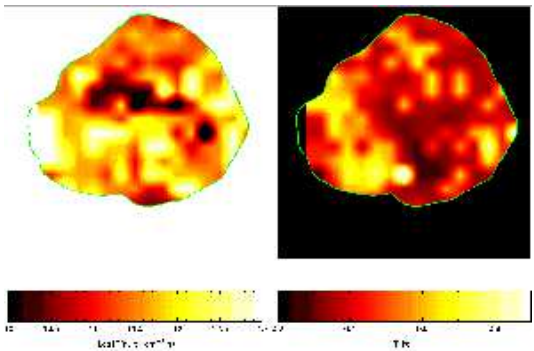}}
}
\caption{Spectral fit parameters for the cool component. Left image: ionisation 
age, Right image: temperature.}
\label{fig5}
\end{figure}

The temperature distribution of the 2 NEI components are similar (but not
the same). The temperature for the cool component ranges between 0.3-1 keV,
displayed in figure~\ref{fig5}, and those for the hot component between 2-6 
keV. These values are similar to values obtained by \cite*{jbleeker-B2:vin96} 
for a coarse subdivision of the remnant in five regions.

A map of the ionisation age of the cool component is also given in
figure~\ref{fig5} and shows a large spread of $10^{10}-10^{13}$cm$^{-3}$s. For
ionisation ages larger than $\sim 10^{12}$ cm$^{-3}$s the plasma is almost in
ionisation equilibrium and cannot be distinguished from equilibrium spectra.
There is also a region of very low ionisation age (less than $3\times 10^{10}$
cm$^{-3}$s) stretching from East to West just above the centre of the remnant.
This region also has a low emissivity and can be understood as a low density
wake just behind and inside of the shocked ejecta. The hot component has a 
more homogeneous ionisation age distribution centered around 
$10^{11}$cm$^{-3}$s (not shown).

\subsection{Elemental abundances}

\begin{figure}[!htb]
\centering
\includegraphics{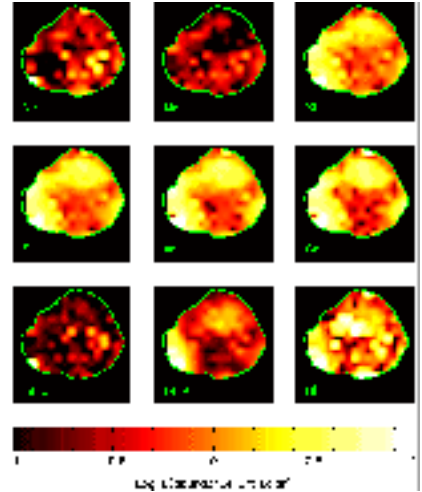}
\caption{Abundance maps for the elements included in the spectral fitting.
All are plotted on the logarithmic scale indicated by the bar at the 
bottom.}
\label{fig6}
\end{figure}

\begin{figure}[!htb]
\centering
\includegraphics{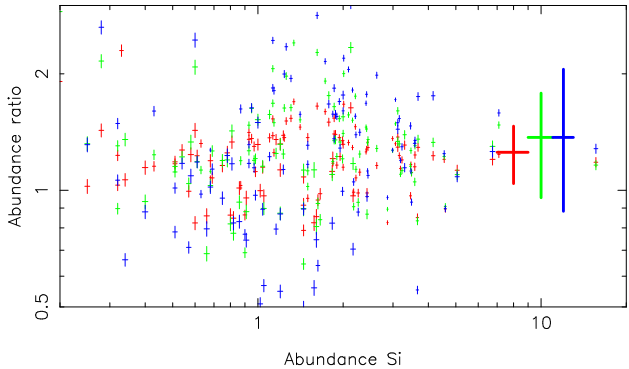}
\caption{The variation in the abundance ratios of S/Si red, Ar/Si green 
and Ca/Si blue as a function of the Si abundance. The large error bars
to the right indicate the mean and rms scatter for the three elements.}
\label{fig7}
\end{figure}

Figure \ref{fig6} is a montage of abundance maps of Ne, Mg, Si, S, Ar, Ca, Fe-L, 
Fe-K and Ni, where the abundance values are with respect to solar. Again we see
considerable variations over the remnant. The Fe-L distribution comes
from the cool component while the Fe-K and Ni are derived
exclusively from the hot component.
The distributions of Si, S, Ar
and Ca, which are all oxygen burning products, are similar and
distinct from carbon burning products, Ne and Mg, and Fe-L.
Figure \ref{fig7} shows the variation in the ratios S/Si, Ar/Si and Ca/Si
with respect to the abundance of Si.
On the one hand these ratios clearly
vary over the remnant but on the other hand, for a Si abundance range spanning
more than two orders of magnitude, these ratios remain remarkably constant.
The thick vertical bars indicate the mean and rms scatter of the
ratio values.

The models for nucleosynthesis yield from massive stars predict that the
mass or abundance ratio $R_{\rm X/Si}$ of ejected mass of any element X
with respect to silicon varies significantly as a function of the
progenitor mass $M$. We show the observed mean values of $R_{\rm X/Si}$
as well as its rms variation in Table \ref{tab1}, together with the predictions
for core collapse SNe models with a progenitor mass of 12, 30 M$_{\sun}$ (ZAMS) 
and for a type Ib model with a  progenitor mass of 60 M$_{\sun}$ (ZAMS).
The model for the 60~\msun\ star included mass loss, in order to mimic
a Type Ib supernova. Abundance values for the 12 and 30 M$_{\sun}$ progenitors 
are taken from 
\cite*{jbleeker-B2:wv95},
while those for the 60 M$_{\sun}$ progenitor are taken from 
\cite*{jbleeker-B2:woo93}.
\begin{table}
\caption[]{Mean measured abundance ratios and rms scatter compared with
theoretical predictions for progenitor masses of 12, 30 and 60 $\rm
M_{\odot}$ (including windloss).}
\centering
\begin{tabular}{|l|ll|lll|}
\hline
ratio & mean & rms & 12 M$_{\sun}$ & 30 M$_{\sun}$ & 60 $\rm 
M_{\odot}$ \\
     \hline
O/Si    & 1.69 & 1.37   & 0.16 & 1.88  &0.71\\
Ne/Si    & 0.24 & 0.37  & 0.12 & 1.24  &0.43\\
Mg/Si    & 0.16 & 0.15  & 0.12 & 2.08  &0.26\\
S/Si    & 1.25 & 0.24  & 1.53 &  0.21  &1.33\\
Ar/Si    & 1.38 & 0.48  & 2.04 & 0.13  &1.02\\
Ca/Si    & 1.46 & 0.68  & 1.62 & 0.11  &1.09\\
FeL/Si    & 0.19 & 0.65  & 0.23 & 0.08 &0.65\\
FeK/Si    & 0.60 & 0.51  & 0.23 & 0.08 &0.65\\
Ni/Si    & 1.67 & 5.52  & 0.68 &  0.29 &    \\
\hline
\end{tabular}
\label{tab1}
\end{table}
The observed abundance ratios for X equal to Ne,
Mg, S, Ar, Ca and Fe-L (related to the iron emissivity of the cool component) 
are consistent with a low progenitor mass of type II 12 M$_{\sun}$ model, but 
the observed  O ratio is an order of magnitude larger than predicted. The type 
II 
30 M$_{\sun}$ model predicts the correct O ratio, but does not produce the 
observed ratios observed in the other elements. The best overall agreement of 
observed 
abundance ratios are provided by the type Ib 60 M$_{\sun}$ model. In all these 
models, most
of the Si, S, Ar and Ca comes from the zones where complete explosive O-burning
and incomplete explosive
Si-burning occurs, and indeed as figure \ref{fig7}
shows these elements track each
other remarkably well.
The correlation between Si and
S is remarkably sharp but not perfect, the scatter in figure \ref{fig7}
is real. These remaining residuals can be attributed to small
inhomogeneities. Table \ref{tab1} also indicate that the rms
scatter of the abundances with respect to Si get larger as Z increases,
S to Ar to Ca and indeed through to Fe and Ni. This is to be expected since
elements close together in Z are produced in the same layers within
the shock collapse structure while those of very different Z are
produced in different layers and at different temperatures.

The Fe which arises from complete and incomplete explosive Si burning, together 
with existing Fe, should give
rise to iron line emission. For both the Fe-L and Fe-K lines we see that
iron abundance varies over the remnant but does not show any
straightforward correlation with either the other elements (there is a very 
large scatter in $R_{\rm Fe/Si}$) or relative to each other.
This is to be expected if most of the iron arises
from complete Si burning. We return to the different morphologies of Si and Fe
later in the discussion.

Ne and Mg are mostly produced in shells where Ne/C burning
occurs, and the relative scatter in terms of $R_{\rm X/Si}$ is indeed
much larger than for S, Ar and Ca (Table \ref{tab1}).
Furthermore the abundance maps of Ne and Mg in figure \ref{fig6} are
mutually similar but very different from the Si, S, Ar and Ca group.

It should be noted that the derivation of O/Si ratio is difficult and less 
reliable than the other ratio's
due to the strong galactic absorption and the poor spectral resolution of the 
EPIC camera's at low energies.
The high resolution XMM RGS maps resolve the O distribution much more clearly 
and show
that this differs from the other elements (\cite{jbleeker-B2:ble01}). We discuss
this briefly in section 5.

\subsection{Dynamics}

\begin{figure}[!htb]
\centering
\includegraphics{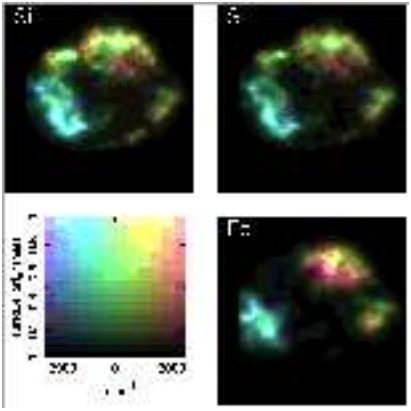}
\caption{Doppler maps derived from Si-K, S-K and Fe-K emission lines.
For each case the surface brightness of the line emission (after
subtraction of the continuum) is shown
colour coded with the Doppler velocity. The coding used
is shown in the bottom left image.}
\label{fig8}
\end{figure}

\begin{figure}[!htb]
\centering
\resizebox{\hsize}{!}{\includegraphics{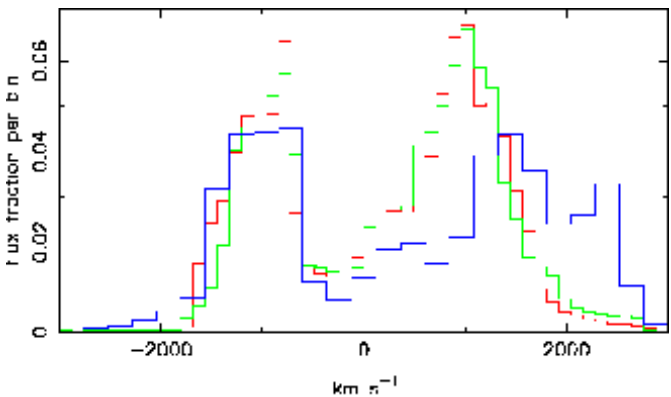}}
\caption{Flux distributions of Si-K (red), S-K (green)
and Fe-K (blue) as a function of measured Doppler velocity.}
\label{fig9}
\end{figure}

It is possible to determine the Doppler shifts of Si-K, S-K and Fe-K since 
these
lines are strong and well resolved. Doppler shifts of these lines have been
calculated by taking advantage of the ionisation age information supplied by 
the
model fits. After fits were made to the full spectrum we froze all the fit 
parameters.
We selected the Si-K (1.72-1.96 keV), S-K (2.29-2.58 keV)
and Fe-K (6.20-6.92 keV) bands for determining their respective
Doppler velocities while ignoring all other line emission.
We then do a fit to each line separately by starting from
the full fit model parameters as a template and subsequently allowing only the
redshift and the abundance of the relevant element to vary.
This method provides a fine tuning of the redshift which in turn
gives the Doppler velocity of the element under scrutiny. Figure \ref{fig8} 
shows the resulting line flux images colour
coded with the Doppler velocity.
The bottom left image is the colour coding used.
The Doppler shifts seen in different areas of the remnant are very
similar in the three lines. The knots in the South East are blue
shifted and the knots in the North are red shifted.
This is consistent with previous measurements (\cite{jbleeker-B2:mar83},
\cite{jbleeker-B2:hol94}, \cite{jbleeker-B2:vin96}).
Moving from large radii towards the centre the shift generally gets larger
as expected in projection. This is particularly pronounced in the North.
At the outer edges the knots are stationary or slightly blue shifted.
Moving South a region of red shift is reached indicating these inner
knots are on the far side of the remnant moving away from us.
The distributions of flux as a function of Doppler velocity are
shown in figure \ref{fig9}. The distributions for Si-K and S-K are very
similar. The Fe-K clearly has a slightly broader distribution for the
red shifted (+ve) velocities. This plot was quite sensitive to
small systematic changes in temperature, ionisation age or abundances
in the spectral fitting since these can potentially have a
profound effect on the derived Doppler velocities. This emphasises the point 
that detailed modeling is an essential prerequisite to arrive at proper 
conclusions.

The X-ray knots of Cas A form a ring because the emitting
plasma is confined to an irregular shell. \cite*{jbleeker-B2:wil02} calculated 
a best fit centre to this ring looking for the position that gave the
most strongly peaked radial brightness distribution (minimum
rms scatter of flux about the mean radius). The best centre
for the combined Si-K, S-K and Fe-K line image was 13 arc second West
and 11 arc seconds North of the image centre (the central Chandra point 
source).
Using this centre the peak flux occurred at a radius of 102 arc secs,
the mean radius was 97 arc secs and the rms scatter about the
mean radius was 24 arc seconds.

\begin{figure}[!htb]
\centering
\includegraphics{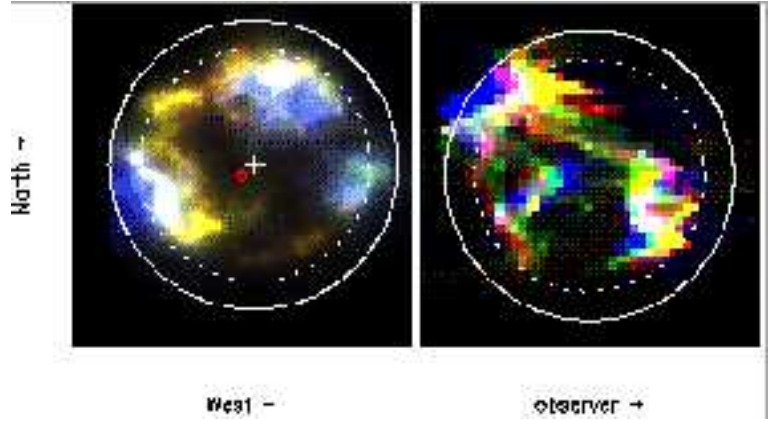}
\caption{The left-hand panel is an image of Si-K (red), S-K (green)
and Fe-K (blue).
The small red circle indicates the position of the
Chandra point source. The white cross
is the best fit centre from the fitting of the radial distribution.
The right-hand panel is a reprojection of the same line fluxes
onto a plane containing the line of sight, North up, observer to the right.
In both panels the outer solid circle is the shock radius
$\rm r_{s}=153$ arc seconds and
the inner circle is the mean radius of the Si-K and S-K flux.}
\label{fig10}
\end{figure}

The left-hand panel of figure \ref{fig10} is a composite image of the
remnant seen in the Si-K, S-K and Fe-K emission lines.
The solid circle indicates
the $\rm r_{s}=153$ arc seconds and the dashed circle is
the mean radius of the Si-K and S-K flux $\rm r_{m}=121$ arc seconds.
The X-ray image of the remnant provides coordinates x-y
in the plane of the sky.
Using the derived radial velocity field within the remnant
we can use the measured Doppler velocities $\rm v_{z}$ to give us
an estimate of
the z coordinate position of the emitting material along the line of
sight thus giving us an x-y-z coordinate
for the emission line flux in each pixel. Using these coordinates we
can reproject the flux into any plane we choose. The right-hand panel
of figure \ref{fig10} shows such
a projection in a plane containing the line-of-sight, North upwards,
observer to the right.
In this reprojection the line emission from Si-K, S-K and Fe-K are
reasonably well aligned for the main ring of knots. The reprojection
is not perfect because the \textit{EPIC-MOS} cameras are unable to resolve
components which overlap along the line-of-sight and this
produces some ghosting just North of the centre of the remnant.
In the plane of the sky Fe-K emission (blue) is clearly
visible to the East
between the mean radius of the Si+S flux and the shock radius.
Similarly in the reprojection Fe-K emission is seen outside the main
ring in the North away from the observer. The Si+S knot in the South away from
the observer
in the reprojection is formed from low surface brightens emission
in the South West quadrant of the sky image.
The X-ray emitting material is very clumpy within the spherical volume
and is indeed surprisingly well characterised by the doughnut shape
suggested by \cite*{jbleeker-B2:mar83}. However the
distribution is distinctly different to that obtained in similar
3-D studies of the optical knots, Lawrence et al. (1995).

The plane of the sky image in figure \ref{fig10} shows that the Fe-K emission to 
the East is at a radius of
140 arc seconds, near the primary shock, with an implied shock velocity of
$\rm u_{s}$ in the range 3500-4500 km s$^{-1}$. The Fe-K emission at large 
radii is
highly reminiscent of SNR shrapnel discovered by \cite*{jbleeker-B2:asc95} 
around 
the Vela SNR. These are almost certainly bullets of
material which were ejected from the progenitor during the
collapse and subsequent explosion. They would initially be expected to
have a radial velocity less than the blast wave but as the remnant
develops, and the shock wave is slowed by interaction with the
surrounding medium, the bullets would overtake the blast wave
and appear outside the visible shock front as is the case
in Vela. Our analysis of the abundances clearly indicates that
the matter responsible for the X-ray emission is ejecta and this must
have been ablated from the bullets rather than swept up by the shock.

\section{High-resolution spectroscopic data}

\begin{figure}
\resizebox{\hsize}{!}{\includegraphics[angle=-90]{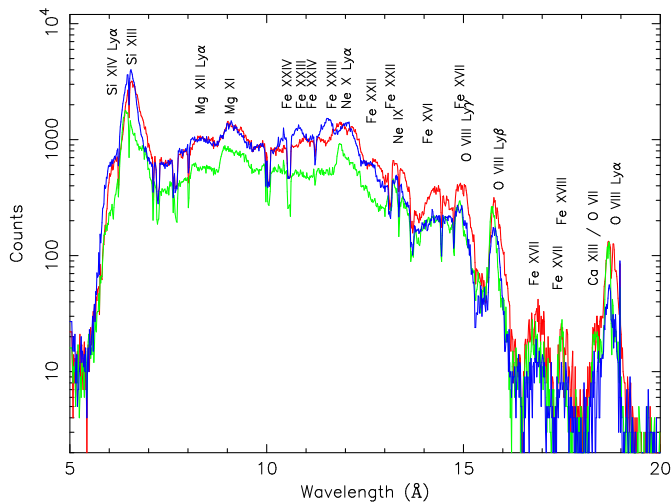}}
\caption[]{First order ($m=-1$) RGS spectra of Cas~A for the three regions. 
Line 
identifications of the principle lines are provided.}
\label{fig11}
\end{figure}

Figure \ref{fig11} displays spectra from three
extraction regions over the
remnant situated in the N, NE and SE of the remnant.  The data from
RGS1 and RGS2 have been combined in order to increase the statistical
weight of the observation.  Several lines of highly ionized species
of Si, Mg, Ne, Fe L and O are
detected in the spectrum.  Due to interstellar absorption no features can be
measured long-ward of $\sim$20~\AA.  The analysis
and interpretation of these spectra
will be the subject of a forthcoming paper.

As a first result \cite*{jbleeker-B2:ble01} extracted images of the 
\ion{O}{viii}
Ly-$\alpha$ and Ly-$\beta$ lines
to probe small scale variations in absorption effects over this
part of the remnant and to investigate the potential presence of
resonance scattering in the limb brightened
shells viewed edge-on. The temperature range relevant for Cas~A
has no influence on the Ly-$\alpha$/Ly-$\beta$ ratio.
Since RGS is a slitless spectrometer it is
possible to extract dispersed monochromatic images of Cas~A.  These images were
converted from wavelength to spatial coordinates using the equation
$\Delta\lambda = d\sin(\alpha)\Delta\phi$, where $\Delta\phi$ is the offset
along the dispersion direction, $\Delta\lambda$  the wavelength shift,
$\alpha$ the angle of incidence on the gratings and $d$ the line distance
of the gratings.  However, any Doppler broadening is also convolved along the
dispersion direction and, depending on its magnitude, Doppler broadening could
distort the RGS images.

\begin{figure}
\resizebox{\hsize}{!}{\includegraphics{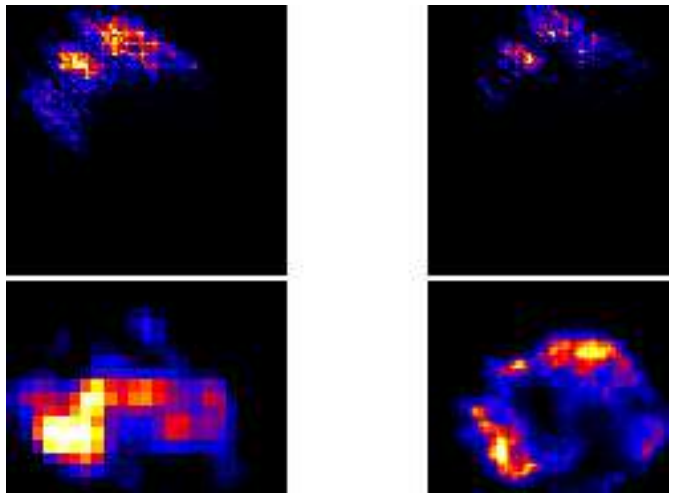}}
\caption[]{RGS images of \ion{O}{viii} Ly-$\beta$ (top left)
and Ly-$\alpha$ (top right), the dispersion direction
runs from top-left to lower-right at 45 degrees with the vertical.
Emission is seen in three blobs 
in the E and N rim. The ratio map of the Ly-$\beta$ to
Ly-$\alpha$ images is shown at the bottom left panel. 
The enhanced spot on the SE rim of the image indicates
a Ly-$\alpha$ emission deficit in this region. For comparison
we show a soft band \textit{EPIC-MOS} image (0.5$-$1.2~keV) in the lower right 
panel.}
\label{fig12}
\end{figure}

The \ion{O}{viii} Ly-$\alpha$ and Ly-$\beta$ maps are presented in
figure~\ref{fig12}. The oxygen emission is seen to originate from three
blobs on the E and N rim of the remnant.  
The emission is faintest in the
SE blob where there also seems to be a Ly-$\alpha$ deficit compared to
Ly-$\beta$. Such a decrease in the 
Ly-$\alpha$/Ly-$\beta$ ratio could possibly be introduced by the presence of
resonance scattering of the \ion{O}{viii} Ly-$\alpha$ photons
in the X-ray bright rims if viewed edge-on (\cite{jbleeker-B2:km95}). 

\section{conclusions}
The hard X-ray continuum maps from XMM-Newton indicate that the
8.0$-$15.0~keV flux, predominantly due to the previously reported high-energy 
spectral tail, does not originate from a few localized regions such as X-ray 
bright knots and filaments, nor does it originate from a limb brightened 
(fractionary) shell
close to the shock front generated by the primary blast wave.  Therefore the 
hard X-ray image does not support in any way the interpretation in terms of 
simple synchrotron emission models. Non-thermal Bremsstrahlung, produced by a 
population of suprathermal electrons, would seem to be a plausible alternative.

The detection of the $^{44}$Sc($^{44}$Ti) emission in the BeppoSax data 
can be reconciled with a progenitor star having a range 30-40 ${\rm M}_{\sun}$, 
assuming a very energetic explosion. It is worth noting, however that the 
relatively high \tiff\ abundance can also be explained by one of the few 
available SN Type Ib models of a star with a ZAMS mass of 60 \msun\ 
(\cite{jbleeker-B2:woo93}). Moreover the elemental abundance ratios predicted by 
this particular model are also roughly consistent with the abundance ratios 
obtained from the XMM-Newton analysis. 

The tight correlation between the variation in abundance of Si, S, Ar, Ca
over an absolute abundance range of two orders of magnitude is
strong evidence for the nucleosynthesis of these ejecta elements by
explosive O-burning and incomplete explosive Si-burning due to the shock
heating of these layers in the core collapse supernova. However the Fe 
emission, both in the Fe-K and the Fe-L
lines, does not show this correlation in any sense. A significant fraction
of the Fe-K emission is seen at larger radii than Si-K and S-K as
convincingly demonstrated in our Doppler derived 3-D reprojection. Moreover the 
Fe-K emission is patchy, reminiscent of large clumps
of ejecta material, rather than shock heated swept up circumstellar material.
The X-ray emission would then arise from ablation of the Fe-ejecta bullets. The 
3-D reprojection also shows a largely bi-polar distribution of the Fe-K 
emission which may indicate that the original explosion was aspherical, 
possibly with axial symmetry.

Future research involves further analysis of spatially resolved high resolution 
spectroscopy as provided by the \textit{RGS}.

\acknowledgement
JV acknowledges support by the NASA through
Chandra Postdoctoral Fellowship Award Number PF0-10011.

\end{document}